\documentclass[english,onecolumn]{IEEEtran}
\usepackage[T1]{fontenc}
\usepackage[latin9]{inputenc}
\usepackage{amsmath}
\usepackage{amsthm}
\usepackage{amssymb}
\usepackage{esint}
\usepackage{color}

\makeatletter
\theoremstyle{plain}
\newtheorem{thm}{\protect\theoremname}
\theoremstyle{plain}
\newtheorem{cor}{\protect\corollaryname}
\theoremstyle{definition}
\newtheorem{defn}{\protect\definitionname}
\theoremstyle{plain}
\newtheorem{lem}{\protect\lemmaname}
\theoremstyle{plain}
\newtheorem{prop}{\protect\propositionname}

\usepackage{mathrsfs}

\makeatother

\usepackage{babel}
\providecommand{\corollaryname}{Corollary}
\providecommand{\definitionname}{Definition}
\providecommand{\lemmaname}{Lemma}
\providecommand{\propositionname}{Proposition}
\providecommand{\theoremname}{Theorem}

\newif\iffullpaper
\fullpapertrue

\newif\ifupperbound
\upperboundtrue

\begin{document}


\title{One-Shot Variable-Length Secret Key Agreement Approaching Mutual
Information}

\author{Cheuk Ting Li and Venkat Anantharam\\
EECS, UC Berkeley, Berkeley, CA, USA\\
Email: ctli@berkeley.edu, ananth@eecs.berkeley.edu}

\maketitle
\begin{abstract}
This paper studies an information-theoretic one-shot variable-length
secret key agreement problem with public discussion.
Let $X$ and $Y$ be jointly
distributed random variables, each taking values in some measurable space.
Alice
and Bob observe $X$ and $Y$ respectively,
can communicate interactively through a public noiseless channel,
and want to agree on a key length and a key that is approximately
uniformly distributed over all bit sequences with the agreed key length.
The public discussion is observed by an eavesdropper, Eve.
The key should be approximately independent of the public discussion, conditional on the key length.
We show that the optimal expected key length is close to the mutual
information $I(X;Y)$ within a logarithmic gap. Moreover, an upper bound
and a lower bound on the optimal expected key length can be written down in
terms of $I(X;Y)$ only. This means that the optimal one-shot performance
is always within a small gap of the optimal asymptotic performance
regardless of the distribution of the pair $(X,Y)$. This one-shot result may
find applications in situations where the components of an i.i.d. pair source $(X^{n},Y^{n})$
are observed sequentially and the key is output bit by bit with small
delay, or in situations where the random source is not an i.i.d. or ergodic process.
\end{abstract}


\section{Introduction}

\allowdisplaybreaks

The information theoretic secret key agreement problem in the source model has been considered in the
asymptotic regime with fixed-length keys by Maurer~\cite{maurer1993secret}
and Ahlswede and Csisz{\'a}r~\cite{ahlswede1993common} (see also~\cite{maurer1999unconditionally,maurer2000information,csiszar2004secrecy,gohari2010information}).
The model is that Alice and Bob observe the components of an i.i.d. pair source $(X^{n},Y^{n})$, where $(X_{i},Y_{i})$ are i.i.d. and $n$
tends to infinity. 
Alice and Bob want to agree on a secret key with length
$nR$ bits using interactive noiseless public discussion, such that an eavesdropper,
Eve, observing the public discussion, asymptotically gets zero information about the key. The optimal
key rate is given by the mutual information $I(X;Y)$~\cite{maurer1993secret,ahlswede1993common}.
These previous works also consider the case where Eve  observes side information $Z^{n}$,
but we will not consider side information at Eve in this paper.

An asymptotic model of this sort is not relevant in settings
where the number of samples of the pair source available is limited, or its statistics do not follow
i.i.d. or other structures that allow asymptotic analysis. 
This motivates us to consider a general one-shot setting where Alice and Bob observe the components $X$
and $Y$ respectively of a jointly distributed pair of random variables $(X,Y)$, where each component takes values in some measurable space.
They wish to agree on the longest secret key
possible
using noiseless public discussion, which is also observed by Eve.
For example,
in an Internet of Things (IoT) deployment, one can envision improving
the security of communication by generating secret keys in real time using the techniques in this paper. The jointly distributed
random variables accessible to the two IoT devices creating such a
secret key could be the result of noise associated to prior transmissions
in the network or generated deliberately. For a concrete scenario
of this kind, if Alice and Bob have earlier received a broadcast message
from Charles, they could use the noise from their respective receptions
to then create a secret key for private communication with each
other. If sufficient resources are available, the broadcast from Charles
might even be simply aimed at generating such dependent randomness
at Alice and Bob. The important constraint, in contrast to asymptotic
approaches, is the real time nature required of secret key generation,
since IoT applications are typically delay-constrained.

 Previous one-shot secrecy results usually
work with a fixed-length setting (e.g. \cite{hayashi2016secretkey,yassaee2015one,tyagi2015converses}).
Here we argue that a variable-length setting is more suitable, for
a similar reason as why a variable-length code is more suitable than a
fixed-length code for one-shot compression. Each value of $X$ and
$Y$ may contain a different amount of information. For example, let
the pair $(X,Y)$ be received signals when the same Gaussian signal is sent through
an additive Gaussian noise broadcast channel. If $X$ is large, then
the amplitude of the signal and the signal-to-noise ratio is likely
to be large, allowing Alice and Bob to agree on a longer key. Unlike
the asymptotic setting, we cannot invoke the law of large numbers
to argue that the amount of usable information is close to the average.
If we require the key length to be fixed, we have to reduce the key
length to accommodate the worse values of $X$ and $Y$, leading to
a waste of information. To make use of all the usable information
in $X$ and $Y$, it is natural to consider a variable-length setting
where Alice and Bob agree on the length of the secret key and this length can
adapt to the values of $X$ and $Y$.
\textcolor{black}{We also note that a universal finite blocklength variable-length key agreement problem has been studied in~\cite{tyagi2017universal}. Nevertheless, the variability of key length in their setting comes from the unknown distribution or type of the source sequence, whereas the variability in this paper comes from the one-shot nature of our setting and the variability of the values of $X,Y$ (or their information density).}

In this paper, we show that the optimal expected key length for one-shot
variable-length secret key agreement with public discussion is close
to $I(X;Y)$ within a logarithmic gap. An upper bound and a lower bound on
the optimal expected key length can be stated in terms of $I(X;Y)$
only, meaning that the optimal one-shot expected key length is always
within a small gap of the optimal asymptotic rate $I(X;Y)$, regardless
of the distribution of $(X,Y)$. Such a result is impossible for fixed-length
keys.

In Section \ref{sec:formulation}, we precisely formulate the one-shot variable-length
secret key agreement setting. In Section \ref{sec:mainres}, we establish upper and lower
bounds on the optimal expected key length in terms of $I(X;Y)$. In
Section \ref{sec:concat}, we show that variable-length keys can be
concatenated to form a fixed-length key in the asymptotic regime.
In Section \ref{sec:split}, we show that a variable-length key can
be applied in situations where a task is to be performed multiple times.


\subsection*{Notation}

Throughout this paper, we assume that $\log$ is to base 2 and the entropy
$H$ is in bits. The binary entropy function is denoted by $H_{\mathrm{b}}(p)=-p\log p-(1-p)\log(1-p)$, for $p \in [0,1]$.
We write $\mathbb{Z}$ for the set of integers, 
$\mathbb{Z}_{\ge0}$ for the set of nonnegative integers, and
$\mathbb{Z}_{>0}$ for the set of positive integers.
We use the notation: $X_{a}^{b} :=(X_{a},\ldots,X_{b})$, $X^{n} :=X_{1}^{n}$
and $[a:b]:=[a,b]\cap\mathbb{Z}$, where $:=$ denotes equality by definition. 
We also use $\stackrel{def}{=}$ for equality by definition. We write $\alpha\,\mathrm{mod}\,\beta$
for the remainder of $\alpha$ divided by $\beta>0$; this lies in
the range $[0,\beta)$. We write $\{0,1\}^{*}$ for $\bigcup_{i=0}^{\infty}\{0,1\}^{i}$.
The length of $w\in\{0,1\}^{*}$ is denoted as $|w|$. The concatenation
of $w_{1},w_{2}\in\{0,1\}^{*}$ is written as $w_{1}\Vert w_{2}$.
For discrete $X$, we write the probability mass function as $p_{X}$.
For continuous $X$, we write the probability density function as
$f_{X}$.  The Bernoulli distribution $p_{X}(0)=1-\alpha$, $p_{X}(1)=\alpha$
is denoted as $\mathrm{Bern}(\alpha)$. The uniform distribution over
a finite set $S$ is denoted as $\mathrm{Unif}(S)$. The total variation
distance between probability distributions on the same finite set is denoted as $d_{\mathrm{TV}}(p_{X},q_{X})\in[0,1]$
and equals $\frac{1}{2} \sum_x |p_X(x) - q_X(x)|$.\medskip{}


\section{Problem Formulation\label{sec:formulation}}

Suppose Alice and Bob observe 
$X$ and $Y$ respectively, where 
$(X,Y)$ is jointly distributed, with each component taking values in some measurable space. 
Alice sends the finite random variable $W_{1}$ (the output of a stochastic mapping on $X$) to Bob.
Bob sends the finite random variable $W_{2}$ (the output of a stochastic mapping on $(Y,W_{1})$)
to Alice, and so forth for $W_{3},\ldots,W_{N}$ until they agree
to stop at time $N\in\mathbb{Z}_{\ge0}$ (which can be random). 
Here $N =0$ corresponds to the situation with no public discussion.
The
public discussion $W^{N}$ is also available to Eve. After public
discussion, Alice and Bob agree on a key length $L\in\mathbb{Z}_{\ge0}$
(i.e., $L$ is a deterministic function of $(X,W^{N})$, and a deterministic function of $(Y,W^{N})$).
Alice produces $A\in[1:2^{L}]$, and Bob produces $B\in[1:2^{L}]$.
We want $A=B$ with high probability, $A,B$ close to being uniform
over $[1:2^{L}]$ and close to being independent of Eve's observation
$W^{N}$. This is captured by the following condition on the total
variation distance
\begin{equation}
\sup_{l\in\mathbb{Z}_{\ge0}}d_{\mathrm{TV}}\left(p_{A,B,W^{N}|L=l}\,,\mathrm{U}_{2}[1:2^{l}]\times p_{W^{N}|L=l}\right)\le\epsilon,\label{eq:tv_eps_bd}
\end{equation}
where we write $\mathrm{U}_{2}[1:2^{l}]$ for the distribution $\mathrm{Unif}(\{(a,a):\,a\in[1:2^{l}]\})$.
Here, on the left hand side of
\eqref{eq:tv_eps_bd} the supremum is over $l\in\mathbb{Z}_{\ge0}$ such that $P(L = l) > 0$, and we
call the left hand side the \emph{distance from the ideal distribution}.
It measures the distance between the actual distribution $p_{A,B,W^{N}|L=l}$
and the ideal distribution where the keys are equal, distributed uniformly
over $[1:2^{l}]$ and independent of $W^{N}$, and where we require the distance
to be small for all $l$, not only averaged over $l$, so we can guarantee
the quality of the key for any key length.\footnote{\textcolor{black}{In~\cite{tyagi2017universal}, a variable-length key with a constraint on the average total variation distance over key lengths is studied. Considering the average distance (instead of the maximum distance in our paper) is undesirable because it is possible for Alice and Bob to declare an extremely long key with low probability, which has an arbitrarily small impact on the average distance, but can increase the expected key length arbitrarily.}}
Define the \emph{maximal
expected key length at distance $\epsilon$}, written as $L_{\epsilon}^{*}(X;Y)$,
as the supremum of $\mathbb{E}[L]$ among all schemes satisfying \eqref{eq:tv_eps_bd}.

\medskip{}

To demonstrate the advantage of variable-length keys in one-shot settings,
consider $X\sim\mathrm{Unif}[1:2^{m}]$ independent of $Q\sim\mathrm{Bern}(7/8)$,
and $Y=X$ if $Q=0$, otherwise $Y|\{X=x,Q=1\}\sim\mathrm{Unif}[1:2^{m}]$.
To generate a variable-length key, Alice and Bob can send the first
$t$ bits of $X$ and $Y$ (containing $m$ bits) respectively through
public discussion. If they match, output $L=m-t$ and let $A,B$ be the
remaining bits of $X,Y$ respectively. Otherwise, output $L=0$. Then
we can achieve $\epsilon=7\cdot2^{-t}$, and $\mathbb{E}[L]\ge(m-t)/8$.
On the other hand, for one-shot fixed-length schemes, any $A,B\in\{0,1\}$
generated by Alice and Bob respectively has
\begin{align*}
 & d_{\mathrm{TV}}(p_{A,B,W^{N}},\,\mathrm{Unif}\{(0,0),(1,1)\}\times p_{W^{N}})\\
 & \ge  d_{\mathrm{TV}}(p_{A,B,W^{N}|Q=1},\,\mathrm{Unif}\{(0,0),(1,1)\}\times p_{W^{N}|Q=1})  -  2/8\\
 & =  \sum_{w^{n}} p_{W^{N}|Q = 1}(w^{n})  \sum_{a=0}^{1} \max \left\{ \frac{1}{2}  -  p_{A,B|W^{N} = w^{n} ,Q = 1}(a, a),0 \right\}  - \frac{1}{4}\\
 & \stackrel{(a)}{=}\sum_{w^{n}}\biggl(p_{W^{N}|Q=1}(w^{n})\\
 & \;\cdot  \sum_{a=0}^{1}  \max \left\{  \frac{1}{2} - p_{A|W^{N}=w^{n} ,Q=1}(a)p_{B|W^{N}=w^{n} ,Q=1}(a),0 \right\}  \biggr) - \frac{1}{4}\\
 & \ge\sum_{w^{n}}p_{W^{N}|Q=1}(w^{n})\inf_{\alpha,\beta\in[0,1]}\biggl(\max\left\{ \frac{1}{2}-\alpha\beta,0\right\} \\
 & \;\;\;\;\;+\max\left\{ \frac{1}{2}-(1-\alpha)(1-\beta),0\right\} \biggr)-\frac{1}{4}\\
 & =\sqrt{2}-1-1/4\,\approx\,0.16.
\end{align*}
Here (a) is because conditioned on $Q=1$, $X,Y$ are independent,
and $I(X;Y|W^{N},Q=1)\le I(X;Y|Q=1)=0$ by Lemma 2.2 in~\cite{ahlswede1993common}
since $W^{N}$ is generated by public discussion, and hence $A-W^{N}-B$
forms a Markov chain conditioned on $Q=1$. Further, the last equality
can be obtained by direct minimization. This
means Alice and Bob cannot even generate 1 bit secret keys that are
approximately $\mathrm{Bern}(1/2)$, approximately independent of
$W^{N}$ and agree with high probability.

Moreover, for the case $X=Y$, the expected length of a variable-length
key can be within a logarithmic gap from $H(X)$ (which can be observed
in the entropy model defined in the next section). This is impossible
in general for fixed-length keys due to the nonuniformity of information
in $X$.\medskip{}

\section{Main Results\label{sec:mainres}}

We present our main result, which is a bound on the gap between $L_{\epsilon}^{*}(X;Y)$
and $I(X;Y)$, which can be stated in terms of $\epsilon$ and $I(X;Y)$
only.
\begin{thm}
\label{thm:agreed_i_bd}For any $X,Y$ and $0<\epsilon<1$, if $I(X;Y) < \infty$, we have
\[
L_{\epsilon}^{*}(X;Y)\ge I(X;Y)-3\log(I(X;Y)+1)-2\log\frac{1}{\epsilon}-15,
\]
\[
L_{\epsilon}^{*}(X;Y)\le(1-\epsilon)^{-1}\left(I(X;Y)+\log3+1\right).
\]
If $I(X;Y)=\infty$ and $\epsilon>0$, then $L_{\epsilon}^{*}(X;Y)=\infty$.
\end{thm}
\medskip{}
The following corollary concerns the regimes $\epsilon=(I(X;Y))^{-\lambda}$
and $\epsilon=2^{-\nu I(X;Y)}$, when $I(X;Y) < \infty$.
\begin{cor}
For any $X,Y$, we have (write $I=I(X;Y)$ and assume $I < \infty$):\\
1. If $\lambda\ge1$ and $I\ge2$, then
\[
I-(3+2\lambda)\log(I+1)-15\le L_{I^{-\lambda}}^{*}(X;Y)\le I+8.
\]
2. If $\nu>0$ and $I\ge\nu^{-1}$, then
\[
(1-2\nu)I-3\log(I+1)-15\le L_{2^{-\nu I}}^{*}(X;Y)\le I+\nu^{-1}+6.
\]
\end{cor}
An implication of this corollary is that the performance of one-shot
variable-length key agreement (i.e., $L_{\epsilon}^{*}(X;Y)$), when
$\epsilon=(I(X;Y))^{-\lambda}$ and $\lambda$ is fixed, is always
within a logarithmic gap from the performance of asymptotic key agreement
(i.e., $I(X;Y)$). For example, if the asymptotic key rate 500 bit/symbol
is achievable, then we know that it is possible to generate a one-shot
variable-length key with $\epsilon=1/500$ and expected length $\ge440$
bits.

\textcolor{black}{When applied to i.i.d. $(X^n,Y^n)$ for fixed $\epsilon$, the lower bound in Theorem~\ref{thm:agreed_i_bd} has a $O(\log n)$ gap from $n I(X;Y)$. This is smaller than the $O(\sqrt{n})$ gap in~\cite{hayashi2016secretkey} (for fixed-length keys) due to the inherent advantage of variable-length keys. A similar logarithmic gap also appears in one-shot variable-length channel simulation and source coding results~\cite{posner1971epsilon,harsha2010communication,braverman2014public,sfrl_isit}.
}

\textcolor{black}{
Also note that the multiplicative gap $(1-\epsilon)^{-1}$ in the upper bound in  Theorem~\ref{thm:agreed_i_bd} is necessary. Consider the erasure source $X \sim \mathrm{Unif}[1:2^m]$, $Y=X$ with probability $1-\epsilon$, $Y=e$ with probability $\epsilon$. Then Alice can output $A=X$, and Bob can output $B=Y$ if $Y \neq e$, and output a random $B$ of length $m$ otherwise. The key length is $m$, which has a multiplicative gap from $I(X;Y)=(1-\epsilon )m$.
}

Before we prove the main result, we introduce an abstract setting,
the entropy model, as an approximation of the variable-length key
model. While the entropy model does not have a concrete operational
meaning, it is easier to analyze and is an important step in proving
the main result.
\begin{defn}
[Entropy model] Alice and Bob observe $X$ and $Y$ respectively and engage in public
discussion $W^{N}$ as in the variable-length key model. After public
discussion, Alice and Bob generate $K_{A}\in\mathbb{Z}_{>0}$ and $K_{B}\in\mathbb{Z}_{>0}$
respectively as the ``secret key'' (instead of $A$, $B$). Here $K_{A}$ is a deterministic function of $(X,W^{N})$ and $K_{B}$ is a deterministic function of $(Y,W^{N})$. There
are no independence requirements between $K_{A},K_{B}$ and $W^{N}$.
Define the \emph{maximal coinciding entropy}, written as $\kappa(X;Y)$,
as the supremum of
\[
H_{=}(K_{A};K_{B}|W^{N})\stackrel{def}{=}\mathbb{P}\{K_{A} = K_{B}\}H(K_{A}|W^{N},\,K_{A} = K_{B})
\]
over all schemes.
\end{defn}
\noindent We first show that $L_{\epsilon}$ can be upper and lower
bounded in terms of $\kappa$.
\begin{lem}
\label{lem:kappa_prop2}For any $X,Y$ and $0<\epsilon<1$, if $\kappa(X;Y)  < \infty$, we have
\[
L_{\epsilon}^{*}(X;Y)\ge\kappa(X;Y)-\log\left(\kappa(X;Y)+1\right)-2\log\frac{1}{\epsilon}-7.082,
\]
\[
L_{\epsilon}^{*}(X;Y)\le(1-\epsilon)^{-1}(\kappa(X;Y)+\log3).
\]
If $\kappa(X;Y)=\infty$ and $\epsilon>0$, then $L_{\epsilon}^{*}(X;Y)=\infty$.
\end{lem}
\begin{IEEEproof}
We first consider the case $\kappa(X;Y)<\infty$.
We prove the lower bound and upper bound separately. For the lower
bound, assume Alice and Bob have created $K_{A},K_{B}\in\mathcal{K}$ respectively, where $\mathcal{K} \subset \mathbb{Z}_{>0}$ 
contains
the range of $K_A$ and $K_B$. We may assume without loss of generality that $P(K_A = K_B) > 0$, since otherwise the lower bound 
(with $H_{=}(K_{A};K_{B}|W^{N})$ replacing $\kappa(X;Y)$) is trivially true. We show how Alice and Bob can generate secret keys $A,B$ respectively using $K_{A},K_{B}$
and further public discussion. The main idea is to partition $\mathcal{K}$
into subsets, and have Alice and Bob send which subset $K_{A}$ and
$K_{B}$ are in. If the two subsets match, Alice and Bob output the
indices of $K_{A}$ and $K_{B}$ within that subset. Otherwise they declare
failure (output $L=0$). The purpose of the partition is twofold:
to group $K_{A}$'s and $K_{B}$'s with similar probabilities into
the same subset so the final key is close to being uniform conditioned
on its length, and to detect errors ($K_{A}\neq K_{B}$) by checking
whether $K_{A},K_{B}$ belong to the same subset. Errors are only
penalized slightly in the entropy model (we simply do not count the
entropy when $K_{A}\neq K_{B}$), but are controlled tightly in the
key agreement setting to have a probability bounded by $\epsilon$
(though the probability of failure, i.e. $P(L=0)$, is not bounded by $\epsilon$),
and hence error detection is necessary.
\textcolor{black}{This technique is similar to spectrum slicing~\cite{han2003information,hayashi2016secretkey}, but here we perform the slicing or partition on the tentative keys $K_{A}, K_{B}$ at the last stage of the scheme, whereas in~\cite{hayashi2016secretkey} the slicing is performed on $X$ at the first stage of the scheme.}

For simplicity, we first assume that Alice and Bob have not used any public discussion
yet (the general case will be addressed later). Fix $0<\delta\le1$,
$0<\epsilon<1$. For $k \in \mathcal{K}$ such that $p_{K_{A}|K_{A}=K_{B}}(k) > 0$, let $\ell(k) :=\lfloor-\delta^{-1}\log p_{K_{A}|K_{A}=K_{B}}(k)\rfloor$. For $t$ in the range of $\ell(\cdot)$, let
$\ell^{-1}(t) :=\{k:\,\ell(k)=t\}$. Note that
$|\ell^{-1}(t)| \ge 1$ for all such $t$. Let $|\ell^{-1}(t)|=\sum_{i=1}^{\beta_{t}}2^{\alpha_{t,i}}$
be the binary representation of $|\ell^{-1}(t)|$ for such $t$, where the $\alpha_{t,i}$'s
are sorted in descending order along $i$. We partition $\ell^{-1}(t)$
by first selecting the $2^{\alpha_{t,1}}$ elements 
$k \in \ell^{-1}(t)$ with the
largest $p_{K_{A}|K_{A}=K_{B}}(k)$ and putting them in the first subset,
then the next $2^{\alpha_{t,2}}$ $k$'s and putting them in the second
subset, and so on. Let $\{S_{i}\}$ be the collection of all these
subsets (note that each of them has a size which is a power of $2$) among the partitions of $\ell^{-1}(t)$
for $t$ in the range of $\ell(\cdot)$. For 
$k \in \mathcal{K}$ such that
$p_{K_{A}|K_{A}=K_{B}}(k) > 0$,
let
$c_{S}(k)$ be the index of the $S_{i}$ that
contains $k$ (i.e., $k\in S_{c_{S}(k)}$), and write $\ell(S_{i})$
for $\ell(k)$ where $k\in S_{i}$ (all $k$'s in $S_{i}$ have the
same $\ell(k)$). By the construction of the partition, we have 
\begin{align}
 & \mathbb{E}\left[\log|S_{c_{S}(K_{A})}|\,\Bigl|\,K_{A}=K_{B}\right]\nonumber \\
 & \ge\mathbb{E}\left[\log|\ell^{-1}(\ell(K_{A}))|\,\Bigl|\,K_{A}=K_{B}\right]-2.\label{eq:prop_ssize}
\end{align}
For each $S_{i}$, let
\[
\rho_{i}=\mathbb{P}\{K_{A}=K_{B}\,|\,K_{A},K_{B}\in S_{i}\},
\]
\[
m_{i}=\max\left\{ \lfloor\log(\epsilon\rho_{i}|S_{i}|)\rfloor,\,0\right\} .
\]
Since $p_{K_{A}|K_{A}=K_{B}}(k) > 0$ for each $k \in S_i$, we have $\rho_i>0$. Further partition $S_{i}$ into $\tilde{S}_{i,1},\ldots,\tilde{S}_{i,2^{-m_{i}}|S_{i}|}$
each with size $2^{m_{i}}$. If we select the partition uniformly
at random,
\begin{align*}
 & \mathbb{P}\left\{ K_{A}\neq K_{B},\,\exists j:K_{A},K_{B}\in\tilde{S}_{i,j}\,|\,K_{A},K_{B}\in S_{i}\right\} \\
 & =\sum_{k_{A}\neq k_{B}\in S_{i}}   p_{K_{A}K_{B}|K_{A},K_{B}\in S_{i}}(k_{A},k_{B})\mathbb{P}\left\{ \exists j:k_{A},k_{B}\in\tilde{S}_{i,j}\right\} \\
 & =\sum_{k_{A}\neq k_{B}\in S_{i}}p_{K_{A}K_{B}|K_{A},K_{B}\in S_{i}}(k_{A},k_{B})\cdot\frac{2^{m_{i}}-1}{|S_{i}|-1}\\
 & \le\epsilon\rho_{i},
\end{align*}
where the last line can be obtained by considering the 2 cases of
$m_{i}$. Hence there exists a fixed partition $\tilde{S}_{i,1},\ldots,\tilde{S}_{i,2^{m_{i}}}$
satisfying
\begin{align}
 & \mathbb{P}\{K_{A}\neq K_{B}\,|\,\exists j:K_{A},K_{B}\in\tilde{S}_{i,j}\}\nonumber \\
 & =\mathbb{P}\left\{ K_{A}\neq K_{B}\,\wedge\,\exists j:K_{A},K_{B}\in\tilde{S}_{i,j}\,|\,K_{A},K_{B}\in S_{i}\right\} \nonumber \\
 & \;\;\;\;\;/\,\mathbb{P}\left\{ \exists j:K_{A},K_{B}\in\tilde{S}_{i,j}\,|\,K_{A},K_{B}\in S_{i}\right\} \nonumber \\
 & \le\epsilon\rho_{i}/\rho_{i}\,=\,\epsilon.\label{eq:prop_probdiff}
\end{align}
Let $c_{\tilde{S}}(k)$ be the index
$j$ of $\tilde{S}_{i,j}$ containing $k$ (i.e., $k\in\tilde{S}_{c_{S}(k),\,c_{\tilde{S}}(k)}$).
Like $c_{S}(k)$, $c_{\tilde{S}}(k)$
is defined for $k \in \mathcal{K}$ such that $p_{K_{A}|K_{A}=K_{B}}(k) > 0$.

If $c_{S}(K_{A})$ and $c_{\tilde{S}}(K_{A})$ are well defined, Alice sends $W_{1}=(c_{S}(K_{A}),\,c_{\tilde{S}}(K_{A}))$ through
public discussion; otherwise Alice sends a failure symbol.
If $c_{S}(K_{B})$ and $c_{\tilde{S}}(K_{B})$ are well defined, 
Bob sends $W_{2}=(c_{S}(K_{B}),\,c_{\tilde{S}}(K_{B}))$; otherwise
Bob sends a failure symbol.
Declare failure (output $L=0$) if either party sends a failure symbol.
If $W_{1}=W_{2}$, Alice outputs the index $A$ of $K_{A}$ in $\tilde{S}_{c_{S}(K_{A}),\,c_{\tilde{S}}(K_{A})}$
(containing $L=\log|\tilde{S}_{c_{S}(K_{A}),\,c_{\tilde{S}}(K_{A})}|$
bits), and Bob outputs the index $B$ of $K_{B}$ in $\tilde{S}_{c_{S}(K_{B}),\,c_{\tilde{S}}(K_{B})}$.
Declare failure if $W_{1}\neq W_{2}$ (output $L=0$).
\iffullpaper
We have 
\begin{align*}
\mathbb{E}[L] & \ge\mathbb{E}[L\mathbf{1}\{K_{A}=K_{B}\}] =\mathbb{P}\{K_{A}=K_{B}\}\mathbb{E}[L\,|\,K_{A}=K_{B}],
\end{align*}
where
\begin{align}
 & \mathbb{E}[L\,|\,K_{A}=K_{B}]\nonumber \\
 & =\mathbb{E}\left[m_{c_{S}(K_{A})}\,\Bigl|\,K_{A}=K_{B}\right]\nonumber \\
 & \ge\mathbb{E}\left[\log|S_{c_{S}(K_{A})}|-\log\frac{1}{\epsilon\rho_{c_{S}(K_{A})}}-1\,\Bigl|\,K_{A}=K_{B}\right]\nonumber \\
 & \ge\mathbb{E}\left[\log|\ell^{-1}(\ell(K_{A}))|\,\Bigl|\,K_{A}=K_{B}\right]\nonumber \\
 & \;\;\;-\mathbb{E}\left[\log\frac{1}{\rho_{c_{S}(K_{A})}}\,\Bigl|\,K_{A}=K_{B}\right]-\log\frac{1}{\epsilon}-3,\label{eq:prop_elbd}
\end{align}
where the last inequality is by \eqref{eq:prop_ssize}. For the first
term,
\begin{align*}
 & \mathbb{E}\left[\log|\ell^{-1}(\ell(K_{A}))|\,\Bigl|\,K_{A}=K_{B}\right]\\
 & =\sum_{t=0}^{\infty}\mathbb{P}\{\ell(K_{A})=t\,|\,K_{A}=K_{B}\}\log|\ell^{-1}(t)|\\
 & \stackrel{(a)}{\ge}\sum_{t=0}^{\infty}\mathbb{P}\{\ell(K_{A}) = t|K_{A} = K_{B}\}\log\frac{\mathbb{P}\{\ell(K_{A}) = t|K_{A} = K_{B}\}}{2^{-\delta t}}\\
 & =\delta\mathbb{E}[\ell(K_{A})\,|\,K_{A}=K_{B}]-H(\ell(K_{A})\,|\,K_{A}=K_{B})\\
 & \stackrel{(b)}{\ge}\delta\mathbb{E}[\ell(K_{A})\,|\,K_{A}=K_{B}]-\log e\\
 & \;\;\;\;-\log\left(\mathbb{E}[\ell(K_{A})\,|\,K_{A}=K_{B}]+1\right)\\
 & \stackrel{(c)}{\ge}\delta\mathbb{E}\left[-\delta^{-1}\log p_{K_{A}|K_{A}=K_{B}}(K_{A})-1\,|\,K_{A}=K_{B}\right]-\log e\\
 & \;\;\;-\log\left(\mathbb{E}\left[-\delta^{-1}\log p_{K_{A}|K_{A}=K_{B}}(K_{A})\,|\,K_{A} = K_{B}\right]+1\right)\\
 & = H (K_{A}|K_{A} = K_{B}) - \log \left(\delta^{-1} H (K_{A}|K_{A} = K_{B}) + 1\right) - \delta - \log e\\
 & \stackrel{(d)}{\ge}H(K_{A}|K_{A}=K_{B})-\log\left(H(K_{A}|K_{A}=K_{B})+1\right)\\
 & \;\;\;\;-\log\frac{1}{\delta}-\delta-\log e,
\end{align*}
where (a) and (c) are because $2^{-\delta(i+1)}\le p_{K_{A}|K_{A}=K_{B}}(k)\le2^{-\delta i}$
for all $k\in\ell^{-1}(i)$, (b) is due to $H(J)\le\log(\mathbb{E}[J]+1)+\mathbb{E}[J]\log(1+1/\mathbb{E}[J])\le\log(\mathbb{E}[J]+1)+\log e$
for any random variable $J\in\mathbb{Z}_{\ge0}$ since the geometric distribution
maximizes the entropy for a given mean, and (d) is because $\delta\le1$.
For the second term,
\begin{align*}
 & \mathbb{E}\left[\log\frac{1}{\rho_{c_{S}(K_{A})}}\,\Bigl|\,K_{A}=K_{B}\right]\\
 & =\sum_{i}\mathbb{P}\{c_{S}(K_{A})=c_{S}(K_{B})=i\,|\,K_{A}=K_{B}\}\log\frac{1}{\rho_{i}}\\
 & =\sum_{i}\frac{\mathbb{P}\{c_{S}(K_{A})=c_{S}(K_{B})=i\}}{\mathbb{P}\{K_{A}=K_{B}\}}\rho_{i}\log\frac{1}{\rho_{i}}\\
 & \le\sum_{i}\frac{\mathbb{P}\{c_{S}(K_{A})=c_{S}(K_{B})=i\}}{\mathbb{P}\{K_{A}=K_{B}\}}(e^{-1}\log e)\\
 & \le\frac{e^{-1}\log e}{\mathbb{P}\{K_{A}=K_{B}\}}.
\end{align*}
Substituting back to \eqref{eq:prop_elbd},
\begin{align}
 & \mathbb{E}[L]\nonumber \\
 & \ge\mathbb{P}\{K_{A}=K_{B}\}\nonumber \\
 & \;\;\;\cdot\left(H(K_{A}|K_{A}=K_{B})-\log\left(H(K_{A}|K_{A}=K_{B})+1\right)\right)\nonumber \\
 & \;\;\;-\log\frac{1}{\delta}-\delta-\log\frac{1}{\epsilon}-e^{-1}\log e-\log e-3\nonumber \\
 & \ge\mathbb{P}\{K_{A}=K_{B}\}H(K_{A}|K_{A}=K_{B})\nonumber \\
 & \;\;\;\;-\log\left(\mathbb{P}\{K_{A}=K_{B}\}H(K_{A}|K_{A}=K_{B})+1\right)\nonumber \\
 & \;\;\;\;-\log\frac{1}{\delta}-\delta-\log\frac{1}{\epsilon}-4.974.\label{eq:prop_elbd2}
\end{align}
Note that the length of the key is
$L  =\mathbf{1}\{W_{1}=W_{2}\}m_{c_{S}(K_{A})}.$
Next we analyze the distribution of the key. Fix $i,j$. For any $a\in[1:2^{m_{i}}]$,
\begin{align*}
 & \mathbb{P}\left\{ A=B=a\,|\,K_{A},K_{B}\in\tilde{S}_{i,j}\right\} \\
 & =\mathbb{P}\{K_{A}=K_{B}\,|\,K_{A},K_{B}\in\tilde{S}_{i,j}\}\mathbb{P}\{A=a\,|\,K_{A}=K_{B}\in\tilde{S}_{i,j}\}\\
 & \ge\mathbb{P}\{K_{A}=K_{B}\,|\,K_{A},K_{B}\in\tilde{S}_{i,j}\}\frac{2^{-\delta(\ell(S_{i})+1)}}{2^{-\delta\ell(S_{i})}2^{m_{i}}}\\
 & =\mathbb{P}\{K_{A}=K_{B}\,|\,K_{A},K_{B}\in\tilde{S}_{i,j}\}2^{-m_{i}-\delta},
\end{align*}
since $2^{-\delta(\ell(S_{i})+1)}\le p_{K_{A}|K_{A}=K_{B}}(k)\le2^{-\delta\ell(S_{i})}$
for all $k\in S_{i}$. Write $\mathrm{U}_{2}([1:2^{l}])=\mathrm{Unif}\left(\{(a,a):\,a\in[1:2^{l}]\}\right)$.
\begin{align*}
 & d_{\mathrm{TV}}\bigl(p_{A,B|K_{A},K_{B}\in\tilde{S}_{i,j}}\,,\,\mathrm{U}_{2}([1:2^{m_{i}}])\bigr)\\
 & =\sum_{a=1}^{2^{m_{i}}}\max\left\{ 2^{-m_{i}}-\mathbb{P}\left\{ A=B=a\,|\,K_{A},K_{B}\in\tilde{S}_{i,j}\right\} ,\,0\right\} \\
 & \le\sum_{a=1}^{2^{m_{i}}}\max \left\{ 2^{-m_{i}}-\mathbb{P}\{K_{A} = K_{B}|K_{A},K_{B}\in\tilde{S}_{i,j}\}2^{-m_{i}-\delta},0 \right\} \\
 & =1-2^{-\delta}\mathbb{P}\{K_{A}=K_{B}\,|\,K_{A},K_{B}\in\tilde{S}_{i,j}\}.
\end{align*}
Hence
\begin{align*}
 & d_{\mathrm{TV}}\Bigl(p_{A,B,W^{2}|W_{1}=W_{2},\,c_{S}(K_{A})=i},\\
 & \;\;\;\;\;\;\mathrm{U}_{2}([1:2^{m_{i}}])\times p_{W^{2}|W_{1}=W_{2},\,c_{S}(K_{A})=i}\Bigr)\\
 & =\sum_{j=1}^{2^{-m_{i}}|S_{i}|}\biggl(p_{W^2|W_{1}=W_{2},\,c_{S}(K_{A})=i}((i,j),(i,j)) \\
 & \;\;\;\;\;\;\;\;\cdot d_{\mathrm{TV}}\bigl(p_{A,B|W_{1}=W_{2}=(i,j)}\,,\,\mathrm{U}_{2}([1:2^{m_{i}}])\bigr)\biggr)\\
 & =\sum_{j=1}^{2^{-m_{i}}|S_{i}|}\biggl(\mathbb{P}\left\{ K_{A}\in\tilde{S}_{i,j}\,|\,\exists j':K_{A},K_{B}\in\tilde{S}_{i,j'}\right\} \\
 & \;\;\;\;\;\;\;\;\cdot d_{\mathrm{TV}}\bigl(p_{A,B|K_{A},K_{B}\in\tilde{S}_{i,j}}\,,\,\mathrm{U}_{2}([1:2^{m_{i}}])\bigr)\biggr)\\
 & \le\sum_{j=1}^{2^{-m_{i}}|S_{i}|}\biggl(\mathbb{P}\left\{ K_{A}\in\tilde{S}_{i,j}\,|\,\exists j':K_{A},K_{B}\in\tilde{S}_{i,j'}\right\} \\
 & \;\;\;\;\;\;\;\;\cdot\left(1-2^{-\delta}\mathbb{P}\{K_{A}=K_{B}\,|\,K_{A},K_{B}\in\tilde{S}_{i,j}\}\right)\biggr)\\
 & =1 -  2^{-\delta}  \sum_{j=1}^{2^{-m_{i}}|S_{i}|}  \mathbb{P} \left\{ K_{A} = K_{B} \in \tilde{S}_{i,j}\,|\,\exists j'  :K_{A},K_{B}\in\tilde{S}_{i,j'}\right\} \\
 & \le1-2^{-\delta}(1-\epsilon),
\end{align*}
where the last inequality is by \eqref{eq:prop_probdiff}. For $l\ge1$,
\begin{align*}
 & d_{\mathrm{TV}}\left(p_{A,B,W^{2}|L=l}\,,\,\mathrm{U}_{2}([1:2^{l}])\times p_{W^{2}|L=l}\right)\\
 & =\sum_{i}\biggl(\mathbb{P}\{c_{S}(K_{A})=i\,|\,W_{1}=W_{2},\,m_{c_{S}(K_{A})}=l\}\\
 & \;\;\;\;\;\cdot d_{\mathrm{TV}}\Bigl(p_{A,B,W^{2}|W_{1}=W_{2},\,c_{S}(K_{A})=i},\\
 & \;\;\;\;\;\;\;\;\;\;\;\;\;\mathrm{U}_{2}([1:2^{m_{i}}])\times p_{W^{2}|W_{1}=W_{2},\,c_{S}(K_{A})=i}\Bigr)\biggr)\\
 & \le1-2^{-\delta}(1-\epsilon)\\
 & \le1-(1-\delta/\log e)(1-\epsilon)\\
 & \le\epsilon+\delta/\log e.
\end{align*}
For any $0<\epsilon'<1$, let $\epsilon=(3/5)\epsilon'$, $\delta=(2/5)\epsilon'\log e$,
then $d_{\mathrm{TV}}(p_{A,B,W^{2}|L=l}\,,\,\mathrm{U}_{2}([1:2^{l}])\times p_{W^{2}|L=l})\le\epsilon'$,
and by \eqref{eq:prop_elbd2},
\begin{align}
\mathbb{E}[L]\ge & \mathbb{P}\{K_{A}=K_{B}\}H(K_{A}|K_{A}=K_{B})\nonumber \\
 & -\log\left(\mathbb{P}\{K_{A}=K_{B}\}H(K_{A}|K_{A}=K_{B})+1\right)\nonumber \\
 & -2\log\frac{1}{\epsilon'}-7.082.\label{eq:prop_elbd3}
\end{align}
\fi
The case $\kappa(X;Y)=\infty$ can be handled by considering a sequence of schemes with $H_{=}(K_{A};K_{B}|W^{N})$ finite and tending to infinity.

\iffullpaper
For the case where Alice and Bob have already used some public discussion
$W^{N}$ to generate $K_{A},K_{B}$, we apply the same arguments for
$p_{K_{A},K_{B}|W^{N}=w^{n}}$ for each $w^{n}$. The additional public
discussion is appended at the end of $W^{N}$ so the overall public discussion
is $W^{N+2}$. We still have $d_{\mathrm{TV}}(p_{A,B,W^{N+2}|L=l}\,,\,\mathrm{U}_{2}([1:2^{l}])\times p_{W^{N+2}|L=l})\le\epsilon'$
by convexity of $d_{\mathrm{TV}}$. Further, we have
\begin{align*}
 & \mathbb{E}_{w^{n}\sim p_{W^{N}}}   \left[\mathbb{P}\{K_{A} = K_{B}|W^{N}  = w^{n}\}H (K_{A}|W^{N}  = w^{n} ,K_{A} = K_{B}) \right]\\
 & =\mathbb{P}\{K_{A}=K_{B}\}H(K_{A}|W^{N},K_{A}=K_{B}).
\end{align*}
Therefore \eqref{eq:prop_elbd3} still holds after replacing $H(K_{A}|K_{A}=K_{B})$
with $H(K_{A}|W^{N},K_{A}=K_{B})$.\\
\medskip{}
\else
\fi

\ifupperbound
For the upper bound, assume Alice and Bob have $A$ and $B$ respectively.
Let
$C$ satisfy $C|\{L=l,\,W^{N}=w^{n}\}\sim\mathrm{Unif}[1:2^{l}]$
for any $l,w^{n}$.
By the coupling characterization of total variation distance, we have
\begin{align*}
 & \mathbb{P}\{A=B=C\,|\,L=l\}\\
 & =1-d_{\mathrm{TV}}\left(p_{A,B,W^{N}|L=l},\,\mathrm{U}_{2}([1:2^{l}])\times p_{W^{N}|L=l}\right)  \ge1-\epsilon.
\end{align*}
We have
\begin{align*}
 & \kappa(X;Y)\\
 & \ge\mathbb{P}\{A=B\}H(A|W^{N},\,A=B)\\
 & =H(A\mathbf{1}\{A=B\}\,|\,W^{N},\,\mathbf{1}\{A=B\})\\
 & \ge H(A\mathbf{1}\{A=B\},\,\mathbf{1}\{A=B=C\}\,|\,W^{N})\\
 & \;\;\;\;-H(\mathbf{1}\{A=B=C\},\,\mathbf{1}\{A=B\})\\
 & \ge H(C\mathbf{1}\{A=B=C\}\,|\,W^{N})-\log3\\
 & \ge\sum_{l=0}^{\infty}\mathbb{P}\{L=l\}H(C\mathbf{1}\{A=B=C\}\,|\,W^{N},L=l)-\log3\\
 & \stackrel{(a)}{\ge}\sum_{l=0}^{\infty}\mathbb{P}\{L=l\}\cdot l\cdot\mathbb{P}\{A=B=C\,|\,L=l\}-\log3\\
 & \ge(1-\epsilon)\mathbb{E}[L]-\log3,
\end{align*}
where (a) is because $C|\{L=l,\,W^{N}=w^{n}\}\sim\mathrm{Unif}[1:2^{l}]$ and
\begin{align*}
 & H(C\mathbf{1}\{A=B=C\}\,|\,W^{N}=w^{n},L=l)\\
 & \ge - \sum_{c\in[1:2^{l}]}\Bigl(\mathbb{P}\left\{ C\mathbf{1}\{A=B=C\}=c\,|\,W^{N}=w^{n},L=l\right\} \\
 & \;\;\;\;\;\cdot \log\mathbb{P}\left\{ C\mathbf{1}\{A=B=C\}=c\,|\,W^{N}=w^{n},L=l\right\} \Bigr)\\
 & \ge - \sum_{c\in[1:2^{l}]}\mathbb{P}\Bigl(\left\{ C\mathbf{1}\{A=B=C\}=c\,|\,W^{N}=w^{n},L=l\right\} \\
 & \;\;\;\;\;\cdot \log\mathbb{P}\left\{ C=c\,|\,W^{N}=w^{n},L=l\right\} \Bigr)\\
 & =\sum_{c\in[1:2^{l}]}\mathbb{P}\left\{ C\mathbf{1}\{A=B=C\}=c\,|\,W^{N}=w^{n},L=l\right\} \cdot l\\
 & =l\cdot\mathbb{P}\left\{ A=B=C\,|\,W^{N}=w^{n},\hat{L}=l\right\} .
\end{align*}
\else
The proof of the upper bound is in~\cite{seckey_arxiv}.
\fi

\end{IEEEproof}
\medskip{}
Since $L_{\epsilon}^{*}$ can be upper and lower bounded by $\kappa$,
in order to prove Theorem \ref{thm:agreed_i_bd}, we can bound $\kappa$
instead. The following lemma bounds $\kappa$ in terms of $I(X;Y)$.
\begin{lem}
\label{lem:kappa_prop_ilb}For any $X,Y$, if $I(X;Y) < \infty$, we have
\begin{align*}
\kappa(X;Y) & \ge I(X;Y)-2\log(I(X;Y)+1)-7.034,
\end{align*}
\[
\kappa(X;Y)\le I(X;Y)+1.
\]
If $I(X;Y)=\infty$, then $\kappa(X;Y)=\infty$.
\end{lem}
\begin{IEEEproof}
\ifupperbound
Assume that $I(X;Y) < \infty$.
We first prove the upper bound.
\begin{align*}
I(X;Y) & =I(X;Y)-I(X;Y|W^{N})+I(X;Y|W^{N},K_{B})\\
 & \;\;\;\;+I(X;K_{B}|W^{N})-I(X;K_{B}|W^{N},Y)\\
 & \stackrel{(a)}{\ge}I(X;K_{B}|W^{N})\\
 & \stackrel{(b)}{\ge}I(K_{A};K_{B}|W^{N})\\
 & \ge I(K_{A};K_{B}|W^{N},\,\mathbf{1}\{K_{A}=K_{B}\})-1\\
 & \ge\mathbb{P}\{K_{A}=K_{B}\}H(K_{A}|W^{N},\,K_{A}=K_{B})-1,
\end{align*}
where (a) is due to $I(X;Y|W^{N})\le I(X;Y)$ by Lemma 2.2 in~\cite{ahlswede1993common} since $W^{N}$ is generated
by interactive communication, and
the Markov chain $X-(Y,W^{N})-K_{B}$, and (b) is due to the Markov chain
$K_{A}-(X,W^{N})-K_{B}$.\medskip{}
\else
The proof of the upper bound is in~\cite{seckey_arxiv}.
\fi

We now prove the lower bound. The main idea is to transmit $X$ from
Alice to Bob, who has side information $Y$, using interactive communication,
and then use the part of $X$ not leaked by the interactive communication
as the key. While this is similar to Slepian-Wolf coding~\cite{slepianwolf1973a}
studied in a one-shot interactive setting in~\cite{braverman2014information,kozachinskiy2016slepian},
here we are concerned with the leakage of information by the interactive communication,
not the amount of communication. Note that if we use the results in~\cite{braverman2014information,kozachinskiy2016slepian},
we obtain a gap on the order of $\sqrt{H(X|Y)}$ instead of $\log I(X;Y)$,
which is undesirable since $H(X|Y)$ can be much larger than $I(X;Y)$,
or even infinite.

We design a scheme for the entropy model as follows. 
First consider the case where $X\sim\mathrm{Unif}[0,1]$ and $Y$ is discrete and finite.
The general case will be addressed later.
Fix $m\in\mathbb{Z}_{>0}$, $0<\epsilon<1/2$.
Alice generates $\tilde{X}_{2},\ldots,\tilde{X}_{2^{m}}\stackrel{iid}{\sim}\mathrm{Unif}[0,1]$. Let
$S_{1} :=\{X,\tilde{X}_{2},\ldots,\tilde{X}_{2^{m}}\}$. At time $i$,
Alice sends $S_{i}$ (as an unordered set, or equivalently a sorted
sequence, of size $2^{m-i+1}$) through public discussion, then Bob finds $\hat{X}_{i}=\arg\max_{x\in S_{i}}f_{X|Y}(x|Y)$.
If 
$f_{X|Y}(\hat{X}_{i}|Y)/\sum_{x\in S_{i}}f_{X|Y}(x|Y)\ge1-\epsilon,$
then Bob declares through public discussion to stop, and Alice produces
$K_{A}\in[1:2^{m-i+1}]$ as the rank of $X$ in $S_{i}$, Bob produces
$K_{B}$ as the rank of $\hat{X}_{i}$ in $S_{i}$. Otherwise, Bob
declares through public discussion to continue, Alice selects $S_{i+1}\subseteq S_{i}$
uniformly among all subsets with size $2^{m-(i+1)+1}$ that contain
$X$, and continues to time $i+1$. The scheme will continue up to
at most time $m+1$ (at which only $S_{m+1}=\{X\}$ is left).
While in this scheme the variable $S_1$, which is part of the public discussion, is not finite (the other $S_i$'s can be transmitted as indices with reference to $S_1$ and are therefore finite), we will later see that it can also be reduced to a finite discrete random variable.

We now analyze the scheme. Let the time at which Bob declares to stop
be $T$. Note that the posterior probability of $\{X=x\}$ (where
$x\in S_{i}$) at Bob at time $i$ is $f_{X|Y}(x|Y)/\sum_{x'\in S_{i}}f_{X|Y}(x'|Y)$.
The posterior error probability is always less than or equal to $\epsilon$
when Bob declares to stop. Hence $\mathbb{P}\{K_{A}\neq K_{B}\,|\,Y=y,\,T=t,\,S^{t}=s^{t}\}\le\epsilon$
for any $y,t,s^{t}$, implying $\mathbb{P}\{K_{A}\neq K_{B}\}\le\epsilon$
and $\mathbb{P}\{K_{A}\neq K_{B}\,|\,Y\}\le\epsilon$ almost surely. 
Let $Q\sim\mathrm{Unif}[0,1]$, independent of all random variables
defined before. Define the event
\[
E=\left\{ K_{A}\neq K_{B}\,\mathrm{or}\,Q\le1-\frac{1-\epsilon}{1-\mathbb{P}\{K_{A}\neq K_{B}\,|\,Y\}}\right\} ,
\]
then $\mathbb{P}\{E|Y\}=\epsilon$ almost surely, and 
\begin{align}
p_{\mathrm{c}}(X,Y) & \stackrel{def}{=}\mathbb{P}\{E^{\mathrm{c}}|X,Y\}\label{eq:sch_pecxy}\\
 & =\mathbb{P}\{K_{A}=K_{B}\,|\,X,Y\}\cdot\frac{1-\epsilon}{1-\mathbb{P}\{K_{A}\neq K_{B}\,|\,Y\}}.\nonumber 
\end{align}

Condition on the event $\{X=x,Y=y\}$ from now on until otherwise
stated. Let $\gamma=f_{X|Y}(x|y)$. Assume Alice continues to generate
$S_{i}$'s after time $T$. Let $S_{m+1}=\{\bar{X}_{1}\}$, $S_{m}=\{\bar{X}_{1},\bar{X}_{2}\}$,
$S_{m-1}=\{\bar{X}_{1},\ldots,\bar{X}_{4}\}$, ..., $S_{1}=\{\bar{X}_{1},\ldots,\bar{X}_{2^{m}}\}$.
Then $\bar{X}_{1}=x$ and $\bar{X}_{2},\ldots,\bar{X}_{2^{m}}\stackrel{iid}{\sim}\mathrm{Unif}[0,1]$.
We also define $\bar{X}_{2^{m}+1},\bar{X}_{2^{m}+2},\ldots$ so that
$\bar{X}_{2},\bar{X}_{3},\ldots\stackrel{iid}{\sim}\mathrm{Unif}[0,1]$.
Assume $\bar{X}_{2}^{\infty}$ is independent of $Q$. Let $V_{i}=f_{X|Y}(\bar{X}_{i}|y)$. Note that $V_i$ has expectation $1$
(conditioned on $\{X=x,Y=y\}$).
Let 
\[
R=\min\Biggl\{ i:\,\sum_{j=2}^{i+1}V_{j}>\frac{\gamma\epsilon}{1-\epsilon}\Biggr\},
\]
then $T\le m+1-\min\{\lfloor\log R\rfloor,\,m\}$ (since by that time
we have $f_{X|Y}(x|y)/\sum_{x'\in S_{i}}f_{X|Y}(x'|y)\ge1-\epsilon$).
For $\alpha<1$, by Markov inequality,
\begin{align*}
\mathbb{P}\left\{ R\le\frac{\alpha\gamma\epsilon}{1-\epsilon}\right\}  & =\mathbb{P}\Biggl\{\sum_{j=2}^{\left\lfloor \frac{\alpha\gamma\epsilon}{1-\epsilon}\right\rfloor +1}V_{j}>\frac{\gamma\epsilon}{1-\epsilon}\Biggr\}\le\alpha.
\end{align*}
Hence,
\begin{align*}
 & \mathbb{E}[m-T+1\,|\,E^{\mathrm{c}}]\\
 & \ge\mathbb{E}\left[\min\{\lfloor\log R\rfloor,\,m\}\,|\,E^{\mathrm{c}}\right]\\
 & =\sum_{i=1}^{m}\mathbb{P}\left\{ \lfloor\log R\rfloor\ge i\,|\,E^{\mathrm{c}}\right\} \\
 & \ge\sum_{i=1}^{m}\max\left\{ 1-\frac{\mathbb{P}\left\{ R<2^{i}\right\} }{\mathbb{P}(E^{\mathrm{c}})},\,0\right\} \\
 & \ge\sum_{i=1}^{m}\max\left\{ 1-2^{i}\frac{1-\epsilon}{\gamma\epsilon\mathbb{P}(E^{\mathrm{c}})},\,0\right\} \\
 & \ge\int_{1}^{m+1}\max\left\{ 1-2^{t}\frac{1-\epsilon}{\gamma\epsilon\mathbb{P}(E^{\mathrm{c}})},\,0\right\} dt\\
 & \ge  \mathbf{1} \left\{ \frac{1-\epsilon}{\gamma\epsilon\mathbb{P}(E^{\mathrm{c}})} \le \frac{1}{2}\right\}  \left( 2\frac{1-\epsilon}{\gamma\epsilon\mathbb{P}(E^{\mathrm{c}})}\log e - \log\frac{1-\epsilon}{\gamma\epsilon\mathbb{P}(E^{\mathrm{c}})} - 1 - \log e \right)\\
 & \;\;\;\;-\max\left\{ -\log\left(\frac{1-\epsilon}{\gamma\epsilon\mathbb{P}(E^{\mathrm{c}})}\right)-(m+1),\,0\right\} \\
 & \ge-\log\frac{1-\epsilon}{\gamma\epsilon\mathbb{P}(E^{\mathrm{c}})}-1-\log e\\
 & \;\;\;\;-\max\left\{ -\log\left(\frac{1-\epsilon}{\gamma\epsilon\mathbb{P}(E^{\mathrm{c}})}\right)-(m+1),\,0\right\} .
\end{align*}
Also note that $K_{A}\neq K_{B}$ if and only if there exists $t\in[1:m+1]$,
$i\in[2:2^{m+1-t}]$ such that
$V_{i}\ge(1-\epsilon)\sum_{j=1}^{2^{m+1-t}}V_{j},$
which is equivalent to
\[
V_{i}\ge\frac{1-\epsilon}{\epsilon}\Biggl(\gamma+\sum_{j\in[2:2^{m+1-t}]\backslash\{i\}}V_{j}\Biggr).
\]
Hence $\mathbb{P}\{K_{A}=K_{B}|X=x,Y=y\}$ only depends on $y$ and
$\gamma=f_{X|Y}(x|y)$, and is nondecreasing in $\gamma$ for fixed
$y$. By \eqref{eq:sch_pecxy}, $p_{\mathrm{c}}(x,y)=\mathbb{P}\{E^{\mathrm{c}}|X=x,Y=y\}$
is nondecreasing in $\mathbb{P}\{K_{A}=K_{B}|X=x,Y=y\}$ for fixed
$y$, and therefore is nondecreasing in $\gamma$ for fixed $y$.

We now remove the conditioning on $\{X=x,Y=y\}$.
\begin{align}
 & \mathbb{E}[m-T+1\,|\,E^{\mathrm{c}}]\nonumber \\
 & \ge\mathbb{E}\left[-\log\frac{1-\epsilon}{f_{X|Y}(X|Y)\epsilon p_{\mathrm{c}}(X,Y)}\,\Bigl|\,E^{\mathrm{c}}\right]-1-\log e\nonumber \\
 & \;\;\;-\mathbb{E}\left[\max \left\{ -\log\frac{1-\epsilon}{f_{X|Y}(X|Y)\epsilon p_{\mathrm{c}}(X,Y)}-m-1,0\right\} \,\Bigl|E^{\mathrm{c}}\right]\nonumber \\
 & \ge \mathbb{E} \left[\log \frac{f_{X|Y}(X|Y)p_{\mathrm{c}}(X,Y)}{1-\epsilon}\Bigl|E^{\mathrm{c}}\right] + \log\epsilon - \frac{\delta_{\epsilon,m}}{1 - \epsilon} - 1 - \log e,\label{eq:sch_emt}
\end{align}
where 
\begin{align*}
\delta_{\epsilon,m} & \stackrel{def}{=}\mathbb{E}\left[\max\left\{ -\log\frac{1-\epsilon}{f_{X|Y}(X|Y)\epsilon}-(m+1),\,0\right\} \right]\\
 & =I(X;Y)+\log\frac{\epsilon}{1-\epsilon}\\
 & \;\;\;\;-\mathbb{E}\left[\min\left\{ -\log\frac{1-\epsilon}{f_{X|Y}(X|Y)\epsilon},\,m+1\right\} \right],
\end{align*}
which tends to 0 as $m\to\infty$ by Fatou's lemma. For the other
term, since $\mathbb{P}\{E^{\mathrm{c}}|Y\}=1-\epsilon$, $E^{\mathrm{c}}$
is independent of $Y$, and 
\begin{align*}
 & \mathbb{E}\left[\log\frac{f_{X|Y}(X|Y)p_{\mathrm{c}}(X,Y)}{1-\epsilon}\,\Bigl|\,E^{\mathrm{c}}\right]\\
 & = \int  \int_{0}^{1} \frac{f_{X|Y}(x|y)p_{\mathrm{c}}(x,y)}{1-\epsilon}\log\frac{f_{X|Y}(x|y)p_{\mathrm{c}}(x,y)}{1-\epsilon}dxdP_{Y}(y).
\end{align*}
Fix $y$. Let $G_{y}=\{x\in[0,1]:\,p_{\mathrm{c}}(x,y)\le1-\epsilon\}$,
$G_{y}^{\mathrm{c}}=[0,1]\backslash G_{y}$. Since $p_{\mathrm{c}}(x,y)$
is nondecreasing in $f_{X|Y}(x|y)$, we have $f_{X|Y}(x_{1}|y)\le f_{X|Y}(x_{2}|y)$
for any $x_{1}\in G_{y}$, $x_{2}\in G_{y}^{\mathrm{c}}$. Let $\ell(t)=t\log t$,
then $\ell'(t)=\log t+\log e$ is increasing,
\begin{align*}
 & \int_{0}^{1}\ell\left(\frac{f_{X|Y}(x|y)p_{\mathrm{c}}(x,y)}{1-\epsilon}\right)dx-\int_{0}^{1}\ell(f_{X|Y}(x|y))dx\\
 & \stackrel{(a)}{=}-\int_{G_{y}}\int_{f_{X|Y}(x|y)p_{\mathrm{c}}(x,y)/(1-\epsilon)}^{f_{X|Y}(x|y)}\ell'(t)dtdx\\
 & \;\;\;\;+\int_{G_{y}^{\mathrm{c}}}\int_{f_{X|Y}(x|y)}^{f_{X|Y}(x|y)p_{\mathrm{c}}(x,y)/(1-\epsilon)}\ell'(t)dtdx\\
 & \ge0,
\end{align*}
since $\ell'(t)$ is increasing, all the $t$'s in the negative integral
in (a) is not greater than the $t$'s in the positive integral in
(a), and 
\begin{align*}
 & \int_{G_{y}^{\mathrm{c}}} \int_{f_{X|Y}(x|y)}^{f_{X|Y}(x|y)p_{\mathrm{c}}(x,y)/(1-\epsilon)}     dtdx- \int_{G_{y}} \int_{f_{X|Y}(x|y)p_{\mathrm{c}}(x,y)/(1-\epsilon)}^{f_{X|Y}(x|y)}      dtdx\\
 & =\int_{0}^{1}\left(\frac{f_{X|Y}(x|y)p_{\mathrm{c}}(x,y)}{1-\epsilon}-f_{X|Y}(x|y)\right)dx\\
 & =\frac{\mathbb{P}\{E^{\mathrm{c}}|Y=y\}}{1-\epsilon}-1 \,=\, 0.
\end{align*}
Hence
\begin{align*}
 &  \int  \int_{0}^{1}\ell\left(\frac{f_{X|Y}(x|y)p_{\mathrm{c}}(x,y)}{1-\epsilon}\right)dxdP_{Y}(y)\\
 & \ge\int\int_{0}^{1}\ell(f_{X|Y}(x|y))dxdP_{Y}(y)\;=\;I(X;Y).
\end{align*}
Substituting back to \eqref{eq:sch_emt},
\begin{align}
 & \mathbb{E}[m-T+1\,|\,E^{\mathrm{c}}]\nonumber \\
 & \ge I(X;Y)+\log\epsilon-(1-\epsilon)^{-1}\delta_{\epsilon,m}-1-\log e.\label{eq:sch_ixy}
\end{align}
Assume Alice selects $S_{i}$ in the following way: Alice observes
$X$, generates $\tilde{X}_{2},\ldots,\tilde{X}_{2^{m}}\stackrel{iid}{\sim}\mathrm{Unif}[0,1]$,
and $S_{1}=\{X,\tilde{X}_{2},\ldots,\tilde{X}_{2^{m}}\}$ (let $\tilde{X}_{1}=X$).
Alice generates a permutation $\Phi$ over $[1:2^{m}]$ uniformly
at random. At time $i$, Alice selects $S_{i}=\{\tilde{X}_{j}:\,\Phi(j)\equiv\Phi(1)\,(\mathrm{mod}\,2^{i-1})\}$.
It is straightforward to check that $S_{i+1}$ is distributed uniformly
among all subsets of $S_{i}$ with size $2^{m-(i+1)+1}$ that contains
$X$. Hence we can assume $S_{i}$'s are generated this way. 
\begin{align}
 & H(K_{A}|T,S^{T},\Phi,E^{\mathrm{c}})\nonumber \\
 & =H(X|T,S^{T},\Phi,E^{\mathrm{c}})\nonumber \\
 & =\mathbb{E}_{t\sim p_{T|E^{\mathrm{c}}}}\biggl[H(X|\,S_{1},\Phi,\,E^{\mathrm{c}},\,T=t)\nonumber \\
 & \;\;\;\;\;\;\;-\sum_{i=2}^{t}I(X;\,S_{i}\,|\,S^{i-1},\Phi,E^{\mathrm{c}},\,T=t)\biggr]\nonumber \\
 & =H(X|\,S_{1},\Phi,T,\,E^{\mathrm{c}})\nonumber \\
 & \;\;\;-\sum_{i=2}^{\infty}\mathbb{P}\{T\ge i\,|\,E^{\mathrm{c}}\}I(X;\,S_{i}\,|\,S^{i-1},\Phi,T,\,E^{\mathrm{c}},\,T\ge i)\nonumber \\
 & \ge H(X|\,S_{1},\Phi,E^{\mathrm{c}})-H(T|E^{\mathrm{c}})-\sum_{i=2}^{\infty}\mathbb{P}\{T\ge i\,|\,E^{\mathrm{c}}\},\label{eq:sch_kappa}
\end{align}
where the last inequality is because $S_{i}$ only has two possibilities
given $S_{i-1}$ and $\Phi$ (depending on whether $\Phi(1)\,\mathrm{mod}\,2^{i-1}=\Phi(1)\,\mathrm{mod}\,2^{i-2}$
or $(\Phi(1)\,\mathrm{mod}\,2^{i-2})+2^{i-2}$). For the first term,
\begin{align*}
 & H(X|\,S_{1},\Phi,E^{\mathrm{c}})\\
 & =\mathbb{E} \left[\sum_{x\in S_{1}} \mathbb{P}\{X=x|S_{1},\Phi,E^{\mathrm{c}}\}\log\frac{1}{\mathbb{P}\{X = x|S_{1},\Phi,E^{\mathrm{c}}\}}\,\Bigl|E^{\mathrm{c}}\right]\\
 & \ge\mathbb{E}\left[\sum_{x\in S_{1}}\mathbb{P}\{X=x|S_{1},\Phi,E^{\mathrm{c}}\}\log\frac{\mathbb{P}\{E^{\mathrm{c}}|S_{1},\Phi\}}{\mathbb{P}\{X=x|\,S_{1},\Phi\}}\,\Bigl|E^{\mathrm{c}}\right]\\
 & =\mathbb{E}\left[\sum_{x\in S_{1}}\mathbb{P}\{X=x|\,S_{1},\Phi,E^{\mathrm{c}}\}\log\frac{\mathbb{P}\{E^{\mathrm{c}}|S_{1},\Phi\}}{2^{-m}}\,\Bigl|E^{\mathrm{c}}\right]\\
 & =m+\mathbb{E}\left[\log\mathbb{P}\{E^{\mathrm{c}}|S_{1},\Phi\}\,|\,E^{\mathrm{c}}\right]\\
 & =m+(1-\epsilon)^{-1}\mathbb{E}\left[\mathbb{P}\{E^{\mathrm{c}}|S_{1},\Phi\}\log\mathbb{P}\{E^{\mathrm{c}}|S_{1},\Phi\}\right]\\
 & \ge m-(1-\epsilon)^{-1}e^{-1}\log e\\
 & \ge m-2e^{-1}\log e,
\end{align*}
where the last inequality is by $\epsilon<1/2$. For the second and
third term in \eqref{eq:sch_kappa},
\begin{align}
 & -H(T|E^{\mathrm{c}})-\sum_{i=2}^{\infty}\mathbb{P}\{T\ge i\,|\,E^{\mathrm{c}}\}\nonumber \\
 & =-m-H(T|E^{\mathrm{c}})+\mathbb{E}[m-T+1|E^{\mathrm{c}}]\nonumber \\
 & \stackrel{(a)}{\ge}-m-(\mathbb{E}[m - T + 1|E^{\mathrm{c}}]+1)H_{\mathrm{b}}\left(\frac{1}{\mathbb{E}[m - T + 1|E^{\mathrm{c}}] + 1}\right)\nonumber \\
 & \;\;\;\;+\mathbb{E}[m-T+1|E^{\mathrm{c}}]\nonumber \\
 & \ge-m+\mathbb{E}[m - T + 1|E^{\mathrm{c}}]-\log\left(\mathbb{E}[m - T + 1|E^{\mathrm{c}}]+1\right)-\log e\nonumber \\
 & \stackrel{(b)}{\ge}-m+\max\left\{ I(X;Y)+\log\epsilon-(1 - \epsilon)^{-1}\delta_{\epsilon,m}- 1 - \log e,0\right\} \nonumber \\
 & \;\;\;- \log \left(\max \left\{ I(X;Y) + \log\epsilon - (1 - \epsilon)^{-1}\delta_{\epsilon,m} - 1 - \log e,0\right\}  + 1\right)\nonumber \\
 & \;\;\;-\log e-0.0861\nonumber \\
 & \ge-m+I(X;Y)+\log\epsilon-(1-\epsilon)^{-1}\delta_{\epsilon,m}\nonumber \\
 & \;\;\;- \log \left(\max \left\{ I(X;Y) + \log\epsilon - (1 - \epsilon)^{-1}\delta_{\epsilon,m},0\right\}  + 1\right) - 3.9715.\label{eq:sch_kappa-1}
\end{align}
where (a) is because $H(T|E^{\mathrm{c}}) = H(m-T+1|E^{\mathrm{c}})$ and the geometric distribution maximizes the entropy
of a nonnegative integer-valued random variable with fixed mean, and
(b) is by \eqref{eq:sch_ixy} and that $t\mapsto t-\log(t+1)$ decreases
by at most $0.0861$. Substituting back in \eqref{eq:sch_kappa},
\begin{align*}
 & H(K_{A}|T,S^{T},\Phi,E^{\mathrm{c}})\\
 & \ge I(X;Y)+\log\epsilon-(1-\epsilon)^{-1}\delta_{\epsilon,m}\\
 & \;\;-\log\left(\max\left\{ I(X;Y) + \log\epsilon - (1 - \epsilon)^{-1}\delta_{\epsilon,m},0\right\}  + 1\right)-5.033.
\end{align*}
Recall that $E^{\mathrm{c}}\subseteq\{K_{A}=K_{B}\}$. Hence
\begin{align}
 & H_{=}(K_{A};K_{B}|T,S^{T},\Phi)\nonumber \\
 & =\mathbb{P}\{K_{A}=K_{B}\}H(K_{A}|T,S^{T},\Phi,K_{A}=K_{B})\nonumber \\
 & =H(K_{A}\mathbf{1}\{K_{A}=K_{B}\}\,|\,T,S^{T},\Phi,\,\mathbf{1}\{K_{A}=K_{B}\})\nonumber \\
 & \ge H (K_{A}\mathbf{1}\{K_{A} = K_{B}\}, \mathbf{1}\{E^{\mathrm{c}}\}|T, S^{T} , \Phi, \mathbf{1}\{ K_{A} = K_{B} \}, \mathbf{1}\{ E^{\mathrm{c}}\} ) - 1\nonumber \\
 & \ge H(K_{A}\mathbf{1}\{E^{\mathrm{c}}\}\,|\,T,S^{T},\Phi,\,\mathbf{1}\{K_{A}=K_{B}\},\,\mathbf{1}\{E^{\mathrm{c}}\})-1\nonumber \\
 & =H(K_{A}\mathbf{1}\{E^{\mathrm{c}}\}\,|\,T,S^{T},\Phi,\,\mathbf{1}\{E^{\mathrm{c}}\})-1\nonumber \\
 & =(1-\epsilon)H(K_{A}|T,S^{T},\Phi,E^{\mathrm{c}})-1\nonumber \\
 & \ge(1-\epsilon)\Bigl(I(X;Y)+\log\epsilon-(1-\epsilon)^{-1}\delta_{\epsilon,m}\nonumber \\
 & \;\;-\log \left(\max \left\{ I(X;  Y) + \log\epsilon - (1 - \epsilon)^{-1}\delta_{\epsilon,m},0\right\}  + 1\right) \Bigr) - 6.033.\label{eq:sch_kappa_heq}
\end{align}
Since $\delta_{\epsilon,m}\to0$ as $m\to\infty$, for $m$ large
enough, we have (write $I=I(X;Y)$)
\[
\kappa\ge(1-\epsilon)\left(I+\log\epsilon-\log\left(\max\left\{ I+\log\epsilon,\,0\right\} +1\right)\right)-6.034.
\]
If $I>2$, substitute $\epsilon=I^{-1}$ , we have
\begin{align*}
\kappa & \ge I-\log I-\log\left(I-\log I+1\right)\\
 & \;\;\;\;+\frac{\log I+\log\left(I-\log I+1\right)}{I}-7.034\\
 & \ge I-2\log(I+1)-7.034.
\end{align*}
It can also be checked that the lemma is true when $I\le2$, since
the right hand side is negative.

Next we consider the case where $X,Y\in\mathbb{Z}_{>0}$ are discrete are finite. Let $\hat{X}|\{X=x\}\sim \mathrm{Unif}[F_X(x-1),F_X(x)]$. Then $\hat{X}\sim\mathrm{Unif}[0,1]$ and $I(\hat{X};Y)=I(X;Y)$. We apply the above scheme over $(\hat{X},Y)$. Since the scheme makes no distinction between values of $\hat{x}$ in the same interval $(F_X(x-1),F_X(x)]$ mapped to the same $x$ (they have the same $f_{\hat{X}|Y}(\hat{x}|y)$ for all $y$), to transmit $S_1$ we only need to transmit the sizes $|S_1 \cap (F_X(x-1),F_X(x)]|$, which are finite.

For the general case where each component of the pair $(X,Y)$ lies in a general measurable space, we apply the above scheme over $(g_1(X),g_2(Y))$, where $g_1(X)$ and $g_2(Y)$ are discretized version of $X$ and $Y$ lying in finite sets. Since (see~\cite{gray2011entropy})
\[
I(X;Y)=\sup_{g_1,g_2:\,g_1(\mathcal{X}),g_2(\mathcal{Y})\,\mathrm{finite}} I(g_1(X);g_2(Y)),
\]
the proof is completed by considering a sequence of discretizations approaching the mutual information. This approach also handles the case where $I(X;Y)=\infty$.
\end{IEEEproof}
\medskip{}


We now complete the proof of Theorem \ref{thm:agreed_i_bd}.
\begin{IEEEproof}
[Proof of Theorem \ref{thm:agreed_i_bd}]The upper bound follows
from Lemma \ref{lem:kappa_prop2} and \ref{lem:kappa_prop_ilb}. For
the lower bound, by Lemma \ref{lem:kappa_prop2} and \ref{lem:kappa_prop_ilb},
\begin{align*}
 & L_{\epsilon}^{*}(X;Y)\\
 & \ge\kappa(X;Y)-\log\left(\kappa(X;Y)+1\right)-2\log\frac{1}{\epsilon}-7.082\\
 & \stackrel{(a)}{\ge}I(X;Y)-2\log(I(X;Y)+1)-7.034\\
 & \;\;\;-\log\left(\max\left\{ I(X;Y)-2\log(I(X;Y)+1)-6.238,0\right\} +1\right)\\
 & \;\;\;-2\log\frac{1}{\epsilon}-0.0861-7.082\\
 & \ge I(X;Y)-3\log(I(X;Y)+1)-2\log\frac{1}{\epsilon}-14.2021
\end{align*}

\noindent where (a) is because $t\mapsto t-\log(t+1)$ decreases by at most
$0.0861$.
\end{IEEEproof}

\section{Concatenating Variable-Length Keys\label{sec:concat}}

Consider the situation where Alice and Bob observe the respective coordinates of a random process
$\{(X_{i},Y_{i})\}_{i\in\mathbb{Z}_{>0}}$ sequentially, where we assume that the pairs $(X_i,Y_i)$ are independent over $i$. 
Instead of
grouping the source symbols into large blocks to allow the generation
of fixed-length keys, they may reduce the delay of key generation by generating a
variable-length key upon observing the respective coordinates of each source symbol pair. These variable-length
keys can be concatenated to form a stream of secret key bits that
can be used as soon as they become available.

Suppose we have two independent variable-length keys with expected
lengths $\mathbb{E}[L_{1}]$, $\mathbb{E}[L_{2}]$ and distances from
ideal distributions $\epsilon_{1},\epsilon_{2}$ respectively. Then
we can concatenate them to form a variable-length key with expected
length $\mathbb{E}[L_{3}]=\mathbb{E}[L_{1}]+\mathbb{E}[L_{2}]$ and
distance from ideal distribution $\epsilon_{3}\le\epsilon_{1}+\epsilon_{2}$.
The distance from ideal distribution grows linearly with the number
of variable-length keys concatenated, which prevents us from concatenating
too many keys. Instead of considering the distance from ideal distribution,
we may consider the entropy and bit error probability instead, as
shown below.
\begin{prop}
\label{prop:pe_ent}Let $(A,B)$ be a variable-length key with expected
length $\mathbb{E}[L]$ and distance from ideal distribution $\epsilon$.
Then, for all $l\in\mathbb{Z}_{\ge0}$ and $i\in[1:l]$ , we have
\begin{equation}  \label{eq:first}
\mathbb{P}\left\{ A[i]\neq B[i]\,|\,L=l\right\} \le\epsilon,
\end{equation}
and
\begin{equation}	\label{eq:second}
H(A\,|\,W^{N},L=l),\,H(B\,|\,W^{N},L=l)\ge l\left(1-2\epsilon\right),
\end{equation}
where we write $A[i]$ for the $i$-th bit of $A$, and $W^N$ denotes the public discussion, stopping at the random time $N$. As a result, if
we concatenate two independent keys $(A_{1},B_{1})$, $(A_{2},B_{2})$
with lengths $L_{1},L_{2}$ and public discussions $W_{1}^{N_{1}},W_{2}^{N_{2}}$
respectively, both with distances from ideal distributions bounded
by $\epsilon$, i.e., $A=A_{1}\Vert A_{2}$, $B=B_{1}\Vert B_{2}$,
$L=L_{1}+L_{2}$, then the same guarantees are preserved, i.e., $\mathbb{P}\left\{ A[i]\neq B[i]\,|\,L=l\right\} \le\epsilon$,
which is \eqref{eq:first} for the concatenated key, and
\begin{equation} \label{eq:third}
H(A|\,W_{1}^{N_{1}} ,W_{2}^{N_{2}},L = l),\,H(B|\,W_{1}^{N_{1}} ,W_{2}^{N_{2}},L = l)\ge l\left(1 - 2\epsilon\right).
\end{equation}
\end{prop}
\iffullpaper
\begin{IEEEproof}
It is straightforward to prove \eqref{eq:first}. We first prove \eqref{eq:second}.
Let $g(t)=-t\log t$ for $t\in[0,1]$. Then, by the concavity
of $g$, for any $\gamma\in[0,1]$,
\[
g(t)\ge g(\gamma)\left(1-\frac{\max\{\gamma-t,\,0\}}{\gamma}-\frac{\max\{t-\gamma,\,0\}}{1-\gamma}\right).
\]
For $l\ge1$,
\begin{align*}
 & H(A|\,W^{N}=w^{n},\,L=l)\\
 & =\sum_{a=1}^{2^{l}}g(p_{A|W^{N}=w^{n},L=l}(a))\\
 & \ge\sum_{a=1}^{2^{l}}g(2^{-l})\biggl(1-\frac{\max\{2^{-l}-p_{A|W^{N}=w^{n},L=l}(a),\,0\}}{2^{-l}}\\
 & \;\;\;\;\;\;-\frac{\max\{p_{A|W^{N}=w^{n},L=l}(a)-2^{-l},\,0\}}{1-2^{-l}}\biggr)\\
 & \ge l2^{-l}\biggl(2^{l}-\frac{d_{\mathrm{TV}}(p_{A|W^{N}=w^{n},L=l},\,\mathrm{Unif}[1:2^{l}])}{2^{-l}}\\
 & \;\;\;\;\;\;-\frac{d_{\mathrm{TV}}(p_{A|W^{N}=w^{n},L=l},\,\mathrm{Unif}[1:2^{l}])}{1-2^{-l}}\biggr)\\
 & \ge l\left(1-2d_{\mathrm{TV}}(p_{A|W^{N}=w^{n},L=l},\,\mathrm{Unif}[1:2^{l}])\right).
\end{align*}
Since $d_{\mathrm{TV}}(p_{A,W^{N}|L=l}\,,\,\mathrm{Unif}[1:2^{l}]\times p_{W^{N}|L=l})\le\epsilon$,
we have
\[
H(A|\,W^{N},\,L=l)\ge l\left(1-2\epsilon\right).
\]
Suppose now that we concatenate two independent keys $(A_{1},B_{1})$, $(A_{2},B_{2})$
with lengths $L_{1},L_{2}$ and public discussions $W_{1}^{N_{1}},W_{2}^{N_{2}}$
respectively, both with distances from ideal distributions bounded
by $\epsilon$, i.e., $A=A_{1}\Vert A_{2}$, $B=B_{1}\Vert B_{2}$,
$L=L_{1}+L_{2}$. It is straightforward to prove \eqref{eq:first} for the concatenated key. To prove \eqref{eq:third}, note that
\begin{align*}
 & H(A|\,W_{1}^{N_{1}},W_{2}^{N_{2}},\,L=l)\\
 & \ge H(A|\,W_{1}^{N_{1}},W_{2}^{N_{2}},L_{1},\,L=l)\\
 & =\sum_{t=0}^{l}\mathbb{P}\left\{ L_{1}=t\,|\,L=l\right\} \Bigl(H(A_{1}|\,W_{1}^{N_{1}},\,L_{1}=t)\\
 & \;\;\;\;\;+H(A_{2}|\,W_{2}^{N_{2}},\,L_{2}=l-t)\Bigr)\\
 & \ge\sum_{t=0}^{l}\mathbb{P}\left\{ L_{1}=t\,|\,L=l\right\} \left(t(1-2\epsilon)+(l-t)(1-2\epsilon)\right)\\
 & =l(1-2\epsilon).
\end{align*}
\end{IEEEproof}
\fi

Then we show that it is possible to construct a fixed-length key in
the asymptotic regime using i.i.d. variable-length keys and a simple
outer code. The following proposition shows that the asymptotic fixed-length
result is implied by the one-shot variable-length result (by applying
Theorem \ref{thm:agreed_i_bd} on $X^{t},Y^{t}$, $\epsilon=t^{-2}$,
and taking $t\to\infty$).
\begin{prop}
Fix $R<\mu(1-H_{\mathrm{b}}(2\epsilon))$. For $i\in\mathbb{Z}_{>0}$, let $(A_{i},B_{i})$
be i.i.d. variable-length keys with respective public discussion $W_{i}$ (we
let $W_{i} :=W_{i}^{N_{i}}$ and omit the superscript), expected length
$\mu$ and distance from ideal distribution $\epsilon$. Then we can
construct a sequence of fixed-length keys $\{(K_{A,n},K_{B,n})\}_{n=1}^{\infty}$, where $K_{A,n},K_{B,n}\in[1:2^{\left\lfloor nR\right\rfloor }]$ is generated
using $A^{n},B^{n}$, and possibly using additional public discussion,
where
\[
\lim_{n\to\infty}\mathbb{P}\left\{ K_{A,n}\neq K_{B,n}\right\} =0,
\]
\[
\underset{n\to\infty}{\lim\inf}\frac{1}{n}H(K_{A,n})\ge R-2\epsilon\mu,
\]
and
\[
\underset{n\to\infty}{\lim\sup}\frac{1}{n}I(K_{A,n};\tilde{W}_{n})\le2\epsilon\mu,
\]
where $\tilde{W}_{n}$ denotes all the public discussion used to generate
$K_{A,n},K_{B,n}$ (including the $W_{i}$'s and the additional public
discussion). Similar conditions hold for $K_{B,n}$.
\end{prop}

\iffullpaper
\begin{IEEEproof}
If $\epsilon > 0$ is small enough, then for all $\xi>0$ small enough we have $R<(\mu+\xi)(1-H_{\mathrm{b}}(2(\epsilon+\xi)))$. Fix such a $\xi > 0$
and fix $n$. Let $\tilde{K}_{A}\in\mathbb{F}_{2}^{n(\mu+\xi)}$ be
the first $\lceil n(\mu+\xi)\rceil$ bits of $A_{1}\Vert A_{2}\Vert\cdots\Vert A_{n}$
(append zeroes if there are not enough bits), and similarly define
$\tilde{K}_{B}$. Let $P\in\mathbb{F}_{2}^{(\lceil n(\mu+\xi)\rceil-\left\lfloor nR\right\rfloor )\times\lceil n(\mu+\xi)\rceil}$
be the parity check matrix of a linear code with minimum distance
at least $2(\epsilon+\xi)n(\mu+\xi)$. This is possible by the Gilbert-Varshamov
bound \cite{gilbert1952comparison,varshamov1957estimate} since $\left\lfloor nR\right\rfloor <n(\mu+\xi)(1-H_{\mathrm{b}}(2(\epsilon+\xi)))$.
Alice sends $P\tilde{K}_{A}$ through public discussion, and Bob finds
$\hat{K}$ with the smallest Hamming distance from $\tilde{K}_{B}$
satisfying $P\hat{K}=P\tilde{K}_{A}$. By Proposition \ref{prop:pe_ent}
and law of large numbers, $\mathbb{P}\{|\{i:\,\tilde{K}_{A}[i]\neq\tilde{K}_{B}[i]\}|\le\epsilon n(\mu+\xi)\}\to1$,
and hence the code can correct the error and $\hat{K}=\tilde{K}_{A}$
with probability tending to 1. Alice outputs $K_{A}\in\mathbb{F}_{2}^{\left\lfloor nR\right\rfloor }$,
the coordinates of $\tilde{K}_{A}$ in the affine subspace $\{v:\,Pv=P\tilde{K}_{A}\}$.
Bob outputs $K_{B}$, the coordinates of $\hat{K}$ in the affine
subspace $\{v:\,Pv=P\tilde{K}_{A}\}$ (Alice and Bob agree beforehand on the
same basis of the subspace). Note that the public discussion is $(W_{1},\ldots,W_{n},P\tilde{K}_{A})=(W^{n},P\tilde{K}_{A})$.
We have
\begin{align*}
 & H(K_{A}|W^{n},P\tilde{K}_{A})\\
 & \ge H(K_{A}|W^{n},L^{n},P\tilde{K}_{A})\\
 & \ge\mathbb{P}\left\{ \Bigl|\frac{1}{n}\sum_{i=1}^{n}L_{i}-\mu\Bigl|\le\xi\right\} \\
 & \;\;\;\cdot H\left(K_{A}\,\biggl|\,W^{n},L^{n},P\tilde{K}_{A},\Bigl|\frac{1}{n}\sum_{i=1}^{n}L_{i}-\mu\Bigl|\le\xi\right).
\end{align*}
By law of large numbers, $\mathbb{P}\left\{ \left|(1/n)\sum_{i}L_{i}-\mu\right|\le\xi\right\} \to1$.
We have
\begin{align*}
 & H\left(K_{A}\,\biggl|\,W^{n},L^{n},P\tilde{K}_{A},\Bigl|\frac{1}{n}\sum_{i}L_{i}-\mu\Bigr|\le\xi\right)\\
 & \stackrel{(a)}{=}H\left(A^{n}\,\biggl|\,W^{n},L^{n},P\tilde{K}_{A},\Bigl|\frac{1}{n}\sum L_{i}-\mu\Bigl|\le\xi\right)\\
 & \ge H \left( A^{n}\,\biggl|\,W^{n},L^{n},\Bigl|\frac{1}{n}\sum L_{i}-\mu\Bigl|\le\xi\right) -n(\mu-\xi)+\left\lfloor nR\right\rfloor \\
 & \stackrel{(b)}{\ge}n(\mu-\xi)(1-2\epsilon)-n(\mu-\xi)+\left\lfloor nR\right\rfloor \\
 & =\left\lfloor nR\right\rfloor -2\epsilon n(\mu-\xi)
\end{align*}
where (a) is because $A_{1}\Vert A_{2}\Vert\cdots\Vert A_{n}$ has
length at most $n(\mu+\xi)$ if $\left|(1/n)\sum_{i}L_{i}-\mu\right|\le\xi$,
and $\tilde{K}_{A}$ is a function of $P\tilde{K}_{A}$ and $K_{A}$;
and (b) is by Proposition \ref{prop:pe_ent}. Hence for sufficiently
large $n$,
\[
\frac{1}{n}H(K_{A}|W^{n},P\tilde{K}_{A})\ge R-2\epsilon(\mu-\xi)-\xi.
\]
Since $I(K_{A};W^{n},P\tilde{K}_{A})\le\left\lfloor nR\right\rfloor -H(K_{A}|W^{n},P\tilde{K}_{A})$,
for sufficiently large $n$,
\[
\frac{1}{n}I(K_{A};W^{n},P\tilde{K}_{A})\le2\epsilon(\mu-\xi)+\xi.
\]
The proof is completed by letting $\xi\to0$.
\end{IEEEproof}
\fi

\medskip{}

\section{Splitting a Variable-Length Key\label{sec:split}}

Another way to obtain fixed-length keys from a variable-length secret
key is by splitting the key. Suppose Alice and Bob share a variable-length
key $A,B$ with length $L$. They want to perform a task multiple
times (e.g. communicating an encrypted message), each time requiring
a fixed-length key with length $t$. Alice and Bob can perform the
task $M=\lfloor L/t\rfloor$ times using different segments of $A$
and $B$ (treated as bit sequences) as the keys. Let the segments
be $\tilde{A}^{\lfloor L/t\rfloor}=\tilde{A}_{1},\ldots,\tilde{A}_{\lfloor L/t\rfloor}$
and $\tilde{B}^{\lfloor L/t\rfloor}$, defined similarly. By the definition of variable-length
keys, we have the following secrecy guarantee for any $m$:
\[
d_{\mathrm{TV}}\Bigl(p_{\tilde{A}^{m},\tilde{B}^{m},W^{N}|M=m},\,\mathrm{U}_{2}([1:2^{t}]^{\otimes m})\times p_{W^{N}|M=m}\Bigr)\le\epsilon,
\]
where $\mathrm{U}_{2}([1:2^{t}]^{\otimes m})$ denotes $\mathrm{Unif}\{(a^{m},a^{m}):\,a_{1},\ldots,a_{m}\in[1:2^{t}]\}$.
This means the total variation distance between the actual distribution
and the ideal one (where $\tilde{A}^{m}=\tilde{B}^{m}$, i.i.d. uniform
over $[1:2^{t}]$ independent of $W^{N}$) is bounded by $\epsilon$.
Any event on $\tilde{A}^{m},\tilde{B}^{m},W^{N}$ (e.g. an error event,
Eve correctly guessing some functions of $\tilde{A},\tilde{B}$, etc.)
has a probability within $\epsilon$ from the probability of that
event measured in the ideal distribution (the probability of error,
the advantage of Eve, etc. are bounded by $\epsilon$). Therefore
Alice and Bob can perform the task an expected $\mathbb{E}[M]\ge\mathbb{E}[L]/t-1$
times while guaranteeing the advantage of Eve is bounded by $\epsilon$.

Consider the payoff function $g(\tilde{a},\tilde{b},v)\in[g_{\min},\,g_{\max}]$
(which can be negative), where $\tilde{a},\tilde{b}\in[1:2^{t}]$
are the keys, and $v$ is Eve's action (e.g. Eve's guess of the message).
The total payoff is $\sum_{i=1}^{M}g(\tilde{A}_{i},\tilde{B}_{i},V_{i})$.
To make the secrecy guarantee stronger, we allow for the hypothetical possibility that Eve observes $\tilde{A}_{i},\tilde{B}_{i}$ strictly causally,
i.e., $V_{i}$ can depend on $W^{N},\tilde{A}^{i-1},\tilde{B}^{i-1},V^{i-1}$. This rules out the possibility of simply reusing the same key for each $i$ and provides a stronger guarantee without actually implying that Eve has access to the previous keys (which would result in compromising previous communications).
Let $g^{*}=\inf_{v}\mathbb{E}[g(C,C,v)]$ be the worst-case expected
payoff in the ideal distribution where $C\sim\mathrm{Unif}[1:2^{t}]$
(since Eve's observation $W^{N}$ is independent of $C$ in the ideal
distribution she can only fix her output at some $v$). Assume $g^{*}>0$
(otherwise we cannot have a positive payoff even if we have a perfect
secret key). The expected payoff


\begin{align}
 & \mathbb{E}\left[\sum_{i=1}^{M}g(\tilde{A}_{i},\tilde{B}_{i},V_{i})\right]\nonumber \\
 & \stackrel{(a)}{\ge}\mathbb{E}\Bigl[\mathbb{E}\Bigl[\sum g(C_{i},C_{i},V_{i})\,\Bigl|\,M\Bigr]\nonumber \\
 & \;\;\;\;\;-d_{\mathrm{TV}}\bigl(p_{\tilde{A}^{M},\tilde{B}^{M},W^{N}|M},p_{C^{M},C^{M},W^{N}|M}\bigr)M(g_{\max}-g_{\min})\Bigr]\nonumber \\
 & \stackrel{(b)}{\ge}\mathbb{E}\left[M\left(g^{*}-\epsilon(g_{\max}-g_{\min})\right)\right]\nonumber \\
 & \ge\left(\frac{\mathbb{E}[L]+1}{t}-1\right)\left(g^{*}-\epsilon(g_{\max}-g_{\min})\right),\label{eq:task_payoff}
\end{align}
where in (a), $C_{i}$ are i.i.d. uniform over $[1:2^{t}]$ independent
of $W^{N}$, and we assume $V_{i}|\{W^{N}=w^{n},C^{i-1}=c^{i-1},V^{i-1}=v^{i-1}\}\sim p_{V_{i}|W^{N},\tilde{A}^{i-1},\tilde{B}^{i-1},V^{i-1}}(\cdot|w^{n},c^{i-1},c^{i-1},v^{i-1})$,
and (b) is because $C_{i}$ is independent of $W^{N},C^{i-1},V^{i-1}$
and therefore $C_{i}$ is independent
of $V_{i}$. We can see that this is close to the ideal payoff $\mathbb{E}[L]g^{*}/t$
when $\epsilon$ is small.

\section{Acknowledgements}

The authors acknowledge support from the NSF grants CNS-1527846,
CCF-1618145, the NSF Science \& Technology Center grant CCF-0939370
(Science of Information), and the William and Flora Hewlett Foundation
supported Center for Long Term Cybersecurity at Berkeley.
The authors thank Himanshu Tyagi and Shun Watanabe for their comments on an earlier version posted on Arxiv.


\bibliographystyle{IEEEtran}
\bibliography{ref}

\end{document}